# γ-Ray Pulsars: Beaming Evolution, Statistics and Unidentified *EGRET* Sources

I.-A. Yadigaroglu and Roger W. Romani[1]

*Department of Physics, Stanford University, Stanford, CA 94305-4060*

## ABSTRACT

We compute the variation of the beaming fraction with the efficiency of high energy γ-ray production in the outer gap pulsar model of Romani and Yadigaroglu. This allows us to correct the fluxes observed for pulsars in the *EGRET* band and to derive a simple estimate of the variation of efficiency with age. Integration of this model over the population of young neutron stars gives the expected number of γ-ray pulsars along with their distributions in age and distance. This model also shows that many of the unidentified *EGRET* plane sources should be pulsars, and predicts the γ-ray fluxes of known radio pulsars. The contribution of unresolved pulsars to the background flux in the *EGRET* band is found to be about 5 %.

*Subject headings:* pulsars — gamma rays — statistics

## 1. Introduction

To date, the only identified galactic point sources of high energy γ-rays are young, spin-powered pulsars. Five pulsars have been now detected at energies $E > 100$ MeV by the *EGRET* experiment on *CGRO*, several other sources have plausible pulsar counterparts, and a number of additional pulsars have been detected at lower energies. Of the *EGRET* sources one, Geminga, is not detected as a radio pulsar; conversely several otherwise likely radio objects have not been detected at *EGRET* energies. Further, there are ~ 30 − 40 sources that have been detected by *EGRET* at low galactic latitude (Fichtel *et al.* 1994). It is important to estimate the fraction of these sources that are rotation-powered pulsars similar to the observed objects.

In the outer magnetosphere model of Romani and Yadigaroglu (1995; hereafter RY), we developed a model for the beaming of high energy γ-ray emission that reproduces the observed properties of individual γ-ray pulsars, including high energy pulse profiles, the relation to the radio pulses, and the non-detection of pulsars such as PSR B0656+14. In RY, it was shown that the statistics of γ-ray pulsars and Geminga-like objects were generally in good agreement with the predictions of our γ-ray pulsar model. Our results are also in general agreement with earlier population studies based on extrapolation from the observed pulsars (*cf.* Helfand 1994), but because of our detailed model predictions our inferences can be made much more specific. In this paper we examine the population predictions in detail and compare these results with the observed galactic plane sources.

To this end we derive the evolution of the γ-ray beaming factor and the pulsar luminosity with age. This evolution differs substantially from that of other high-energy pulsar models and shows that the γ-ray pulsars are a biased subset of the young pulsar population. In addition these sums allow us to estimate the mean distances and luminosities of the galactic plane sources. Comparison with the observations shows that most of these are likely to be young pulsars with $10^4 < \tau_c < 10^6$ y. These computations give important guidelines to assist in detecting other young pulsars and provide estimates of the contribution of these objects to the general flux in the galactic plane. Finally, there are other widely studied models for the production of pulsar γ-rays in the literature. The most important of these is the "polar cap" model (*cf.* Harding and Daugherty 1993, Sturner and Dermer 1994). We compute the statistics and expected luminosities of γ-ray pulsars in this model, finding that in the context of the standard picture of radio pulsars these models do not reproduce the properties of the observed population, in contrast to our outer magnetosphere sums.

## 2. Beaming Factor and Efficiency Evolution

In RY we showed that the γ-ray pulses observed from young pulsars can be explained by emission arising from the upper surface of an acceleration gap in the outer magnetosphere, when the full effects of aberration, retarded potential and time-of-flight across the magnetosphere are included. This model is an outgrowth of the "outer gap" picture of Cheng, Ho and Ruderman (1986, CHR) (see also Ho, 1989), but entails the full zone extending from the null charge ($\Omega \cdot \mathbf{B} = 0$) surface to the speed of light

---

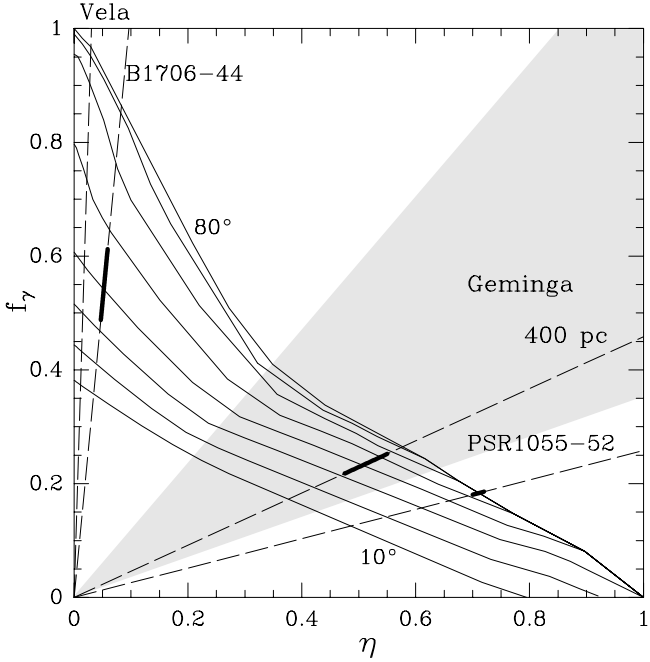

Fig. 1.— The beaming factor $f_\gamma$ as a function of the total gap efficiency $\eta$ (including both photons and pairs) for various $\alpha$. The diagonal lines are combinations of $f_\gamma$ and $\eta_\gamma$ allowed by the measured photon flux for individual pulsars. The $\alpha$ range indicated by polarization and beam shape studies is shown with heavy lines. Geminga's distance is in the range 250 to 500 pc; our efficiency law gives $d \sim 400$ pc; the resulting $\alpha$ agrees with the $\gamma$-ray pulse shape.

cylinder, bounded below by the surface of last closed field lines, and above by field lines a variable distance away. The gap width is specified at the null charge surface and incorporates some rough idealizations of the physical processes limiting gap growth.

In CHR it was argued that as a $\gamma$-ray pulsar ages it will experience greater difficulty in completing the pair production that establishes the upper surface of the gap and the radiation zone. We model here a similar behavior and note that as the gap widens and the distance between the last closed field lines and the radiating surface increases there are two important effects. First, more of the open field lines occupy the gap; this implies that a larger fraction of the spin-down power is expended in accelerating charges at $\sim$ the co-rotation charge density $\rho_{GJ} = 0.07\ B_{12}\ P^{-1}\ cm^{-3}$ in the gap. This power is down-converted and re-radiated at the upper surface of the gap zone (Chiang and Romani 1994) and thus older pulsars with larger gaps will be more efficient at producing $\sim$ GeV $\gamma$-rays (CHR). Second, as the gap widens, the locus of the radiation surface moves closer to the dipole axis and the $\gamma$-ray beam narrows. This results in a smaller beaming fraction with the $\gamma$-rays largely concentrated to the spin equator.

The gap thus spans a variable fraction of the open field lines; with an assumed constant radiation efficiency from the accelerated particles this fraction is in fact $\eta_\gamma = \dot{E}_\gamma/\dot{E}_{SD} = \dot{E}_\gamma/I\Omega\dot{\Omega}$, the $\gamma$-ray efficiency. For each gap width we compute both this efficiency and the fraction of the solid angle $f_\gamma$ swept out on the sky by the resulting $\gamma$-ray beam. The true gap efficiency should include the energy emitted in the relativistic wind of $e^\pm$ pairs; for the Crab this is $\dot{E}_{e^\pm} = \dot{N}_{e^\pm}\gamma m_e c^2 \gtrsim 10^{37-38} s^{-1} \cdot 10^6 (2 \times 511\ keV) \sim 5 \times 10^{37}$ erg s$^{-1} \gg L_\gamma$. So for steep spectrum pulsars with copious pair production such as Crab and to a lesser extent Vela, the true gap width is sufficient to ensure that the real efficiency $\eta$ remains larger than $\sim 10$ %. While this is important to the details of the pulse shape for these very young pulsars, it makes little difference to $f_\gamma$. In Figure 1 we plot $f_\gamma$ against $\eta$ for a range of magnetic inclinations $\alpha$. The decrease in beaming factor with increased efficiency is clear, although previous works (*cf.* Helfand 1994) have assumed that $f_\gamma$ is uncorrelated with $\eta$ and in fact constant.

In RY we showed that the $\alpha$ of the observed $\gamma$-ray pulsars can be estimated from both the $\gamma$-ray and radio data. In addition, the observed *EGRET* $\gamma$-ray fluxes and distances allow us to compute the pulsar luminosity *phase-averaged within the $\gamma$-ray beam*. Since in this model, $\gamma$-ray beams have a roughly constant phase-averaged flux, we see that the total $\gamma$-ray luminosity is

$$L_\gamma = 4\pi d^2 F_\gamma \cdot f_\gamma = \eta_\gamma \cdot \dot{E}_{SD}.$$

Thus the observed fluxes $F_\gamma$, distances and spin-down luminosities define the diagonal lines shown in Figure 1. With the observational estimates of $\alpha$, we can then derive the true $\gamma$-ray efficiencies. These are given in Table 1. While the pulsed luminosity peaks strongly at $\sim$ 1 GeV for the older objects, Crab has a much softer spectrum and it is important to include the X-ray flux in its total emitted power. We estimate that the fraction of power in the GeV band scales as $\tau_4$ for pulsars with a characteristic age less than $10^4 \tau_4$ y. From the observed values for PSR B1706-44 and PSR B1055-52 alone we derive the phenomenological evolution of $\eta_\gamma$ for all high energy pulsed emission, finding

$$\eta_\gamma = 3.2 \times 10^{-5}\ \tau^{0.76} \qquad (1)$$

with characteristic age $\tau$ in years; the results agree well with estimates for the other objects. Note that the beaming corrections result in a significantly flatter evolution curve than the $\tau^{\sim 1.2}$ fitted by Thompson *et al.* (1995, T95). Detailed models of gap closure will be needed to replace this empirical law by the theoretical scaling with spin parameters. The cut-off to this evolution occurs as the beaming factor drops to zero at $\sim 10^6$ y.

We have ranked the known pulsars by estimating their $E > 100$ MeV flux from (1) and their distances, using an average inclination $\alpha = 60°$. The top 15 objects for which $f_\gamma$ is also $> 5$ % are listed in Table 1; five of these are the observed *EGRET* pulsars. There are *EGRET* point sources possibly associated with three of the candidate pulsars. Only pulse upper limits exist for the other objects. In a number of cases, we have estimates of $\alpha$; for these pulsars $F_{mod}$ and $f_\gamma$ have been corrected accordingly. It is interesting to note that for several of the pulsars with upper limits below our predicted flux the known magnetic geometry indicates that the $\gamma$-ray pulse should be beamed away from the Earth. From an examination of the $f_\gamma$, this is likely to be true for most the other sources, as well.

Table 1: $\gamma$-Ray Pulsars and Candidates: Fluxes, Efficiencies and Beaming Fractions

| PSR | $\log \tau(y)$ | $d$(kpc) | $F_\gamma$ [a] | $\alpha(°)$ | $f_\gamma$ | $F_{mod}$ [a,b] | $\eta_\gamma$ [b] | Comments |
|---|---|---|---|---|---|---|---|---|
| B0531+21 | 3.10 | 2.00 | 10. | 80 | 0.86 | 9.8 | 0.0072 | $F_\gamma \sim 110$ including X-ray flux |
| B0833−45 | 4.06 | 0.50 | 71. | 65 | 0.76 | 120. | 0.038 | |
| B1706−44 | 4.24 | 1.82 | 8.3 | 45 | 0.55 | 8.3 | 0.053 | |
| J0630+17 | 5.53 | $\sim$0.40 | 37. | $\sim$40 | 0.23 | 37. | 0.51 | |
| B1055−52 | 5.73 | 1.53 | 4.2 | $\sim$90 | 0.18 | 4.2 | 0.71 | [c] |
| B1951+32 | 5.03 | 2.50 | <1.7 | | 0.53 | 20. | | Wide pulse suggests small $\alpha$ |
| B0535+28 | 5.78 | 1.50 | | | 0.15 | 10. | | $EGRET$ source J0542+26,[d] $F_\gamma \sim 2.8$ |
| B0114+58 | 5.44 | 2.12 | <1.8 | | 0.32 | 5.5 | | |
| B0656+14 | 5.05 | 0.76 | <1.1 | $\sim$10 | 0.25 | 4.8 | | $\beta = +8°$, beamed away |
| B0355+54 | 5.75 | 2.07 | <1.9 | 50 | 0.16 | 4.1 | | $\beta = +4°$, poor $\alpha$ fit, pulsed flux? (T95) |
| B0740−28 | 5.20 | 1.89 | <1.9 | 40 | 0.32 | 2.9 | | $\beta = -10°$, beamed away |
| B1221−63 | 5.84 | 2.29 | <2.2 | 50 | 0.10 | 2.6 | | $\beta = +7°$, beamed away |
| B1634−45 | 5.77 | 3.83 | <3.9 | | 0.15 | 2.2 | | |
| B1823−13 | 4.33 | 4.12 | <3.4 | $\sim$30 | 0.42 | 2.1 | | $EGRET$ source J1823-12,[d] $F_\gamma \sim 9$ |
| B1046−58 | 4.31 | 2.98 | <2.2 | | 0.70 | 1.6 | | $EGRET$ source J1047-58, $F_\gamma \sim 7$ |

[a] $E_\gamma > 100$ MeV $\gamma$-ray flux in units of $10^{-10}(\frac{\text{erg}^2}{\text{cm}^2\,\text{s}})$. For the candidates we convert from $EGRET$ count rate limits on narrow pulses (T95) using $\langle E_\gamma \rangle = 750$ MeV, as appropriate to pulsars in this age range. Three pulsars have plausible associations with $EGRET$ point sources.
[b] Using Eq. 1 and the estimated $\alpha$. For the candidate pulsars with measured radio polarization sweep we use the $\alpha$ value from Rankin (1993). Where none is available we have assumed $\alpha = 60°$.
[c] RY used $\alpha \sim 70$ of Lyne and Manchester (1988), which happened to give the correct $\gamma$-ray pulse width and phase offset for the generic $\tau \sim 10^{4.5}$ gap used. With the evolved gap, $\alpha \sim 90$ (Rankin 1993) is convincing as it gives the observed $\gamma$-ray pulse and radio interpulse.
[d] The $EGRET$ sources consistent with PSRs B0535+28 and B1823-13 may be associated with the supernova remnants G 180.0-1.7 (S147) and G 18.8+0.3 (Kes 67) (Sturner and Dermer, 1995); much of the flux may be due to local cosmic ray enhancement.

### 3. Population Estimates

The beaming factor and efficiency evolution derived above allow us to integrate over the population of young pulsars. We adopt conventional assumptions, namely that pulsars are born at a rate $R \sim 1/100$ y at spin periods of $\sim 10$ ms, that magnetic fields are distributed as a Gaussian in $\log(B)$, with mean $\log(B)=12.7$ and dispersion $\sigma(\log B) = 0.3$, and that the magnetic inclination $\alpha$ is distributed isotropically at birth. We ignore any exponential field decay for these young objects. The evolution of the radio beam is taken from Biggs (1990):

$$W(\text{P}) = 6.2° \, \text{P}^{-1/2}$$

while for the pulsar luminosity, in standard units of mJy kpc$^2$ at 400 MHz, we follow Lorimer *et al.* 1993 and take

$$L_{400} = 2.8 \text{ mJy kpc}^{-2} \, \dot{\epsilon}^{1/2}$$

where $\dot{\epsilon} = \min(\text{P}\dot{\text{P}}/10^{-15}, 10^2)$, with a Gaussian spread in $\log(L_{400})$ of $\sigma_L = 0.8$. These pulsars are distributed in the galaxy with a radial law $\Psi_R(R) \, R \, dR$ that peaks at the $\sim 5$ kpc ring, falling as a Gaussian towards the outer galaxy

$$\Psi_R(R) \propto \begin{cases} \exp(-\frac{(|R|-3.7)^2}{2R_i^2}) \times 0.743, & |R| \leq 3.7 \text{ kpc} \\ \exp(-\frac{R^2}{2R_o^2}), & |R| > 3.7 \text{ kpc} \end{cases}$$

where $R_i$=1.5 kpc, $R_o$=4.8 kpc and the constant serves to match the inner and outer galaxy solutions (Johnston 1994). We also give the pulsars a Gaussian scale height at birth of $z_0 \sim 80$ pc and a galactic $z$ velocity drawn from the 1-D projection of the 2-D Lyne and Lorimer (1994) distribution.

A Monte Carlo integration of the population consists of samples at each Galactic longitude $l$, with draws of the pulsar properties $\alpha$, B, $t$, $z_0$ and $v_z$. We compute the $\gamma$-ray and radio luminosities and beaming factors, determine the fraction of the sky that is swept out by the radio and/or $\gamma$-ray beams to find the detection probability and the maximum detectable distance. Pulsars are then populated at various $d$ in direction $l$ according to the normalized population $\Psi[R(l,d)]$ and then placed in the $b$ bin corresponding to the height $|z| = |z_0 + v_z \cdot t(\tau)|$ to form the Monte Carlo sample. For the radio pulsars, our goal is not to model the observations in detail, so we do not model individual surveys; a flux threshold of 5 mJy approximates the average survey depth for the bulk of the young pulsar population. For the $\gamma$-ray detections, we adopt a flux threshold of $3 \times 10^{-10}$ erg cm$^{-2}$ s$^{-1}$, comparable to the faintest $5\sigma$ sources in the first *EGRET* galactic plane source catalog (Fichtel *et al.* 1994).

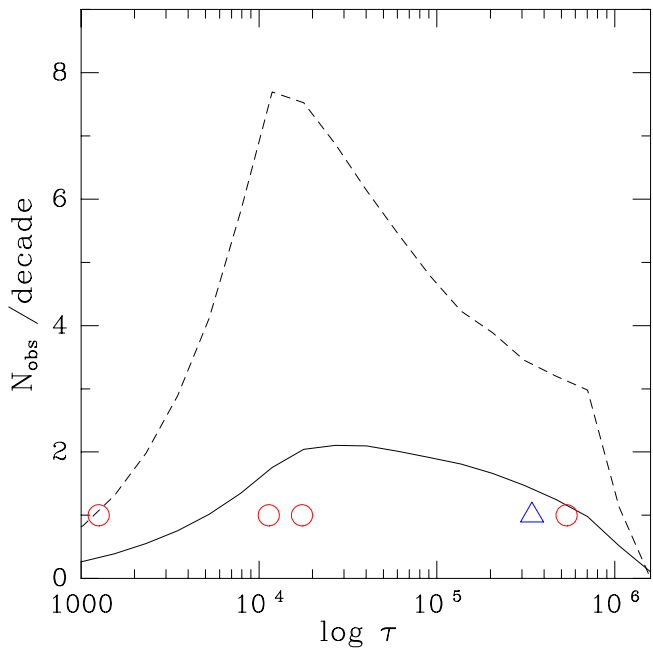

Fig. 2.— The expected number of detectable $\gamma$-radio pulsars (solid line) and Geminga-like pulsars (dashed line) as a function of characteristic age $\tau$. *EGRET* pulsars (open circles) and Geminga (triangle) are also indicated.

Several important features of the populations are explained by our sums. In Figure 2 the detectable numbers of objects, per decade in $\tau$, expected for *EGRET* are displayed as a function of characteristic age. Noting that the total number of radio pulsars expected in the *EGRET* band is 5 in this sum, the agreement with the observed sample (open circles) is quite good. The number of Geminga-like objects not visible in the radio is 17. Only Geminga (triangle) has at present been identified, but similar objects are believed to be present in the unidentified source list. We assess this possibility by comparing with the source catalog in the next section. Very deep radio surveys would increase the fraction of radio-detected pulsars to $\sim 1/2$ in our model. Varying the initial B fields within a reasonable range changes the total number of objects by a factor of order two. The beaming evolution prevents most pulsars as old as $10^6$ y from being seen as $\gamma$-ray objects.

The limited $\gamma$-ray sensitivity assures that the high energy pulsars are a more local sample of the galaxy than the radio objects, as is shown in Figure 3. Our sums also explain that, like the observed radio pulsars, the detectable $\gamma$-ray pulsars are a biased subset of the parent pulsar population (*cf.* Bailes and Kniffen 1992). In particular for a given $\tau$ the observed sample is heavily weighted towards low B (and low P). This is a natural result of our beaming and efficiency evolution, and is clear in the *EGRET* pulsars. However, it should be noted that very low $B$ fields (*e.g.* PSR B1951+32) may not allow efficient particle multiplication and GeV $\gamma$-ray production, while high field objects (*e.g.* PSR B1509-58) may have a strongly enhanced soft synchrotron spectrum; further polarization studies are needed to see if these objects exhibit true spectral variations or primarily beaming effects.

Looking at the five *EGRET* pulsars it can be speculated that some initial B field decay occurs on short time scales. We have run our simulation with a power law B field decay that matches the trend observed for the *EGRET* pulsars. The effect is to increase the number of detections for $\tau$ between $10^5$ and $10^6$ y.

Before we leave the population estimates, we note that we have computed the population models for the luminosity and beaming laws of the polar cap model of Dermer and Sturner (1994). Using their formulation, we find that only 0.04 radio pulsars and 0.03 radio quiet objects should be detected in the present *EGRET* source catalog, and less than 0.01 objects as young as the Crab are expected. An error in Dermer and Sturner's sums was the assumption of a source detection threshold of $10^{-11}$ erg cm$^{-2}$ s$^{-1}$; such fluxes may be reached at $\sim 3\sigma$ at high galactic latitude for long integrations, but are much fainter than the source detection threshold in the galactic plane. Of course, if we assume increased *EGRET* sensitivity, the properties of the detected objects match poorly with those of the observed pulsars in this model; in particular, comparing with Figure 3, the mean source distance of $\sim 10$ kpc is seen to be much larger than that of the observed pulsars. Other polar cap models fare similarly poorly in population sums.

## 4. Unidentified *EGRET* Sources and Conclusions

In the 1$^{\text{st}}$ *EGRET* Galactic plane ($|b| < 10°$) catalog there are 42 sources with detection significance $> 5\sigma$, five of which are the young pulsars mentioned above. Our goal is to estimate what fraction of this sample is due to unidentified young pulsars. Several caveats are in order. Given the extreme sensitivity of the source detection to the background model, it is likely that a modest fraction of these sources are spurious. Conversely several additional steady sources will be found at comparable levels in later catalogs. Present in this sample are certainly some AGN visible through the plane of the Galaxy; extrapolation from high latitude counts suggests that this may contribute $\sim 10 \pm 5$ sources to the plane sample (Kanbach 1995). Additionally, concentrations in the diffuse emission, either from molecular clouds or from SNR enhancements in the local cosmic ray density (Sturner and Dermer 1995) may also be present, if unresolved. Nevertheless, we can use the available source positions and fluxes to check agreement with our model predictions.

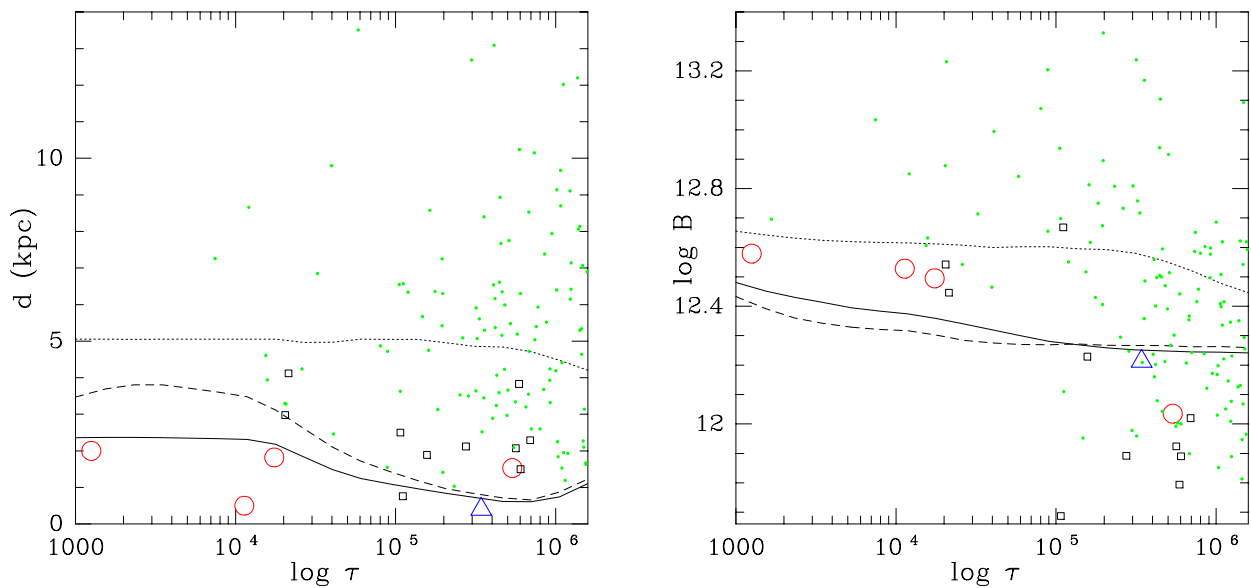

Fig. 3.— The computed median distance and B field for radio pulsars (dotted line), $\gamma$-radio pulsars (solid line) and Geminga-like pulsars (dashed line) as a function of characteristic age $\tau$. Observed objects are also indicated: young radio pulsars (dots), *EGRET* pulsars (open circles), Table 1 candidate pulsars (small squares), and Geminga (triangle).

In Figure 4 we follow Kanbach (1995) in showing the distributions in $l$ and $b$ of the *EGRET* sources (histograms) along with the model predictions for radio-detectable and Geminga-like $\gamma$-ray pulsars. We have assumed a symmetric galaxy and reflected the data about $l = 0$ and $b = 0$. The agreement confirms the average source distance in our sums. We note that there is an excess at low $b$ compared to the model; this is likely due to contamination by "diffuse emission" sources. We have also found a correlation between the mean $b$ angle and age, due both to proper motion and to the decreased sample depth for these fainter objects. Finally, from our sums we also expect one or two unidentified sources to be pulsars with $|b| > 10°$.

What may be done to further test this model? Foremost, spectral data on the unidentified sources will be essential in separating the pulsar candidates (with spectral indices $\sim -2$ to $-1.3$) from the much softer galactic emission. To a lesser extent AGN may be identified by spectral differences, but these sources should be primarily separated by $\gamma$-ray variability (present in virtually all high latitude sources). The remaining hard spectrum, steady sources are good pulsar candidates. A few of these objects should yield counterparts to very deep radio searches. Also several pulsar IDs should be supported by identification of plerionic SNR in the *EGRET* error boxes. To some extent the $b(\tau)$ correlation expected for pulsars may be tested in the unidentified sources by noting that the detected pulsars show some spectral index variation with $\tau$. Duplication of Halpern and Holt's identification of lower-energy (*e.g.* X-ray) pulsations, will prove difficult except for the brightest sources. Scaling from the Crab and PSR B1509-58, ASCA observations for the youngest pulsars may have some success. The sample of both Geminga-like objects and $\gamma$-ray/radio pulsars can of course be increased by more careful analysis of the observations, allowing detections to fainter flux levels.

Looking at the candidate pulsars in Table 1, one can see that if further analysis of the increasing *EGRET* photon database can decrease the limiting sensitivity by a factor of $\sim 3-5$, a number of additional radio pulsars should be within reach. Given the dependence of the model flux on $\alpha$ and corrections to the flux from the details of the pulse shape, a number of pulsars ranked below those in Table 1 will also be interesting targets; an additional $\sim 20$ objects have model fluxes within a factor of 3 of that for PSR B1823-13. However, at typical ages of $\log(\tau) > 5.5$ most of these will have small $f_\gamma$ and will not be visible from Earth. Since the $\gamma$-ray beams are confined to the spin equator for the efficient older pulsars; we require $\zeta = \alpha + \beta \gtrsim 80°$ to see these pulsars. Objects with interpulses will have large $\alpha$ and are thus excellent targets; PSRs B1719-37, B1736-29, B0906-49 and B1822-09 are likely to be seen with a factor of 5 increase in *EGRET* pulse flux sensitivity. We note also that our beaming model provides expected $\gamma$-ray pulse phases and profiles, which can enhance the significance in these pulse searches.

We can at some level test the significance of the beaming corrections from Table 1. For $\tau \sim 10^3$y pulsars, $f_\gamma$ suggests that $\sim 80\%$ should be detected; 1/1 are seen. For $\tau \sim 10^4$y pulsars the detection fraction should be $\sim 60\%$. At present 2/4 from Table 1 are detected; we might expect one of the other candidates to be confirmed. For $10^5 - 10^6$ y pulsars, the mean $f_\gamma$ is $\sim 25$ %. Of the 9 objects in the Table, 2 are presently seen (not counting Geminga); we might therefore expect one of the candidates to be confirmed. We have not included an $\alpha$ bias implicit for radio detections in these estimates.

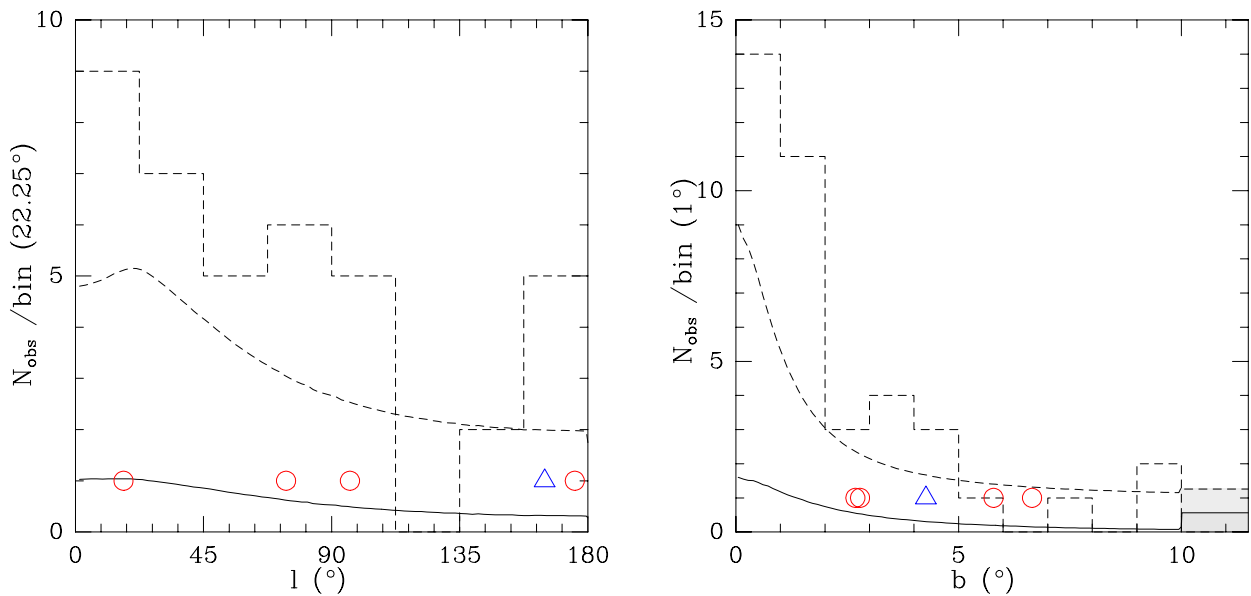

Fig. 4.— The distribution of unidentified *EGRET* galactic plane sources in $l$ and $b$. Curves show model results for $\gamma$-ray pulsars with (solid line) and without (dashed line) detectable radio beams. The histograms of identified *EGRET* sources and the observed pulsars are also shown. A total of 10 AGNs distributed uniformly were added to the dashed curves. The shaded bin shows the total number of expected pulsars for $|b| > 10°$.

Finally, it is traditional to estimate the contribution of the pulsars not visible as discrete sources to the diffuse galactic background (*e.g.* Bailes and Kniffen 1992). In our model, the integrated flux from sources below our identification threshold is $1.9 \times 10^{-8}$ erg cm$^{-2}$ s$^{-1}$. This is only about 5 % of the measured plane flux. To some extent this may be testable since this component should have the $E^{-1.5}$ spectrum and $l$ and $b$ distribution of the $\tau \sim 10^5$ y pulsars.

In conclusion, it is clear that the outer magnetosphere model of RY can provide a good explanation for much of the galactic $\gamma$-ray source population. While the luminosity law is at present phenomenological, the $\eta_\gamma$ and $f_\gamma$ variations in the model are important in interpreting the observations. Further observations and analyses can test this model and more careful studies of the brightest pulsars, coupled with detailed spectral modeling (Chiang and Romani 1994) can illuminate the mechanics of the pulsar magnetosphere itself.

Support for this work was provided by NASA grants NAGW-2963 and NAG5-2037.

**REFERENCES**

Bailes, M. and Kniffen, D.A. 1992, ApJ, 391, 659

Biggs, J.D. 1990, MNRAS, 245, 514

Cheng, K.S., Ho, C., and Ruderman, M. A. 1986, ApJ, 300, 522 (CHR)

Chiang, J. and Romani, R.W. 1994, ApJ, in press

Dermer, C.D. and Sturner, S.J. 1994, ApJ, 420, L75

Fichtel, C.E., *et al.* 1994, ApJ, in press

Halpern, J.P. and Holt, S.S. 1992, Nature, 357, 222

Halpern, J.P. and Ruderman, M. 1993, ApJ, 415, 286

Harding, A.K. and Daugherty, J 1993, in *Isolated Pulsars*, eds K.A. van Riper, R. Epstein, and Ho, C. (New York: Cambridge), 279

Helfand, D.J. 1994, MNRAS, 267, 490

Ho, C. 1989, ApJ, 342, 396

Johnston, S. 1994, MNRAS, 268, 595

Kanbach, G. 1995, preprint

Lorimer, D.R., Bailes, M., Dewey, R.J. and Harrison, P.A. 1993, MNRAS, 263, 403

Lyne, A.G. and Lorimer, D.R., Nature, 369, 127

Rankin, J.M. 1993, ApJ, 405, 285

Romani, R.W. and Yadigaroglu, I.-A. 1995, ApJ, in press (RY)

Sturner, S.J. and Dermer, C.D. 1994, ApJ, 420, L79

Sturner, S.J. and Dermer, C.D. 1995, A&A, in press

Thompson, D.J., *et al.* 1995, ApJ, in press (T95)